# A Bayesian dose-response meta-analysis model: simulation study and application

Tasnim Hamza[1], Andrea Cipriani[2], Toshi A. Furukawa[3], Matthias Egger[1], Nicola Orsini[4], Georgia Salanti[1].

[1]Institute of Social and Preventive Medicine, University of Bern, Bern, Switzerland. [2]Department of Psychiatry, University of Oxford. [3]Department of Health Promotion and Human Behavior, and Department of Clinical Epidemiology, Graduate School of Medicine/School of Public Health, Kyoto University, Kyoto, Japan. [4]Department of Global Public Health, Karolinska Institutet, Stockholm, Sweden.

Abstract

Dose-response models express the effect of different dose or exposure levels on a specific outcome. In meta-analysis, where aggregated-level data is available, dose-response evidence is synthesized using either one-stage or two-stage models in a frequentist setting. We propose a hierarchical dose-response model implemented in a Bayesian framework. We present the model with cubic dose-response shapes for a dichotomous outcome and take into account heterogeneity due to variability in the dose-response shape. We develop our Bayesian model assuming normal or binomial likelihood and accounting for exposures grouped in clusters. We implement these models in R using JAGS and we compare our approach to the one-stage dose-response meta-analysis model in a simulation study. We found that the Bayesian dose-response model with binomial likelihood has slightly lower bias than the Bayesian model with the normal likelihood and the frequentist one-stage model. However, all three models perform very well and give practically identical results. We also re-analyze the data from 60 randomized controlled trials (15,984 participants) examining the efficacy (response) of various doses of antidepressant drugs. All models suggest that the dose-response curve increases between zero dose and 40 mg of fluoxetine-equivalent dose, and thereafter is constant. We draw the same conclusion when we take into account the fact that five different antidepressants have been studied in the included trials. We show that implementation of the hierarchical model in Bayesian framework has similar performance to, but overcomes some of the limitations of the frequentist approaches and offers maximum flexibility to accommodate features of the data.

TH and GS are funded by the European Union's Horizon 2020 research and innovation programme under grant agreement No 825162. ME was supported by special project funding (grant 17481) from the Swiss National Science Foundation.

# 1   Introduction

Dose-outcome associations examine the effect of different levels of exposure (for example, levels of smoking or drug doses) on a health outcome [1, 2]. In pairwise meta-analysis [3–5], combining dose-response associations from different studies and settings may lead to more precise and generalizable conclusions [6]. When aggregate-level data are available from multiple studies, dose-response associations can be synthesized using either a one-stage or two-stage model. The one-stage model is implemented as a linear mixed model, which estimates a dose-response fixed effect and accounts for the heterogeneity by allowing slopes to vary across studies [7]. In a two-stage model, the dose-response model is fitted first within each study, and then the regression coefficients (or shape characteristics) are synthesized across studies [8–10].

The one-stage model takes into account heterogeneity but provides relevant information via the estimate of a between-studies variance-covariance matrix. The two-stage model employs standard meta-analytical techniques and provides the usual heterogeneity measures, such as $I^2$, in case this is of interest. However, to fit non-linear shapes, frequentist implementation of the two-stage model requires multiple dose levels to be reported in each study. For example, if the dose-response curve is assumed to be approximated by a $p$-order polynomial, all studies need to report outcomes for at least $p + 1$ dose levels. This situation will result in excluding studies that report $p$ dose levels or fewer levels.

The one-stage and two-stage models are implemented in a frequentist setting, and their performance has been evaluated in simulations and examples [11]. Fitting dose-response meta-analysis in a Bayesian framework, in the form of a hierarchical model, is, in our view, highly desirable. First, Bayesian models [12, 13] can be easily extended to incorporate, for example, study-specific covariates, to combine observational and randomized data, or to deal with multiple outcomes and exposure types. Second, one can employ informative priors for the dose-response shape to reflect expert knowledge or evidence from external data sources. Third, one can easily extend the model to explore the variation in dose-response curves within and across groups of similar exposures or drugs. Finally, probabilistic statements follow naturally as the posterior distributions can be interpreted as the true distributions of quantities of interest [14, 15].

The paper is structured as follow. In Section 2 we present a Bayesian hierarchical dose-response meta-analysis model with normal or binomial likelihood and the cluster-specific dose-response model. The evaluation of the properties of the models follows in Section 3, alongside comparisons with the frequentist model in a simulations study. In Section 4, we re-analyse a dataset of the dose-response association of various doses of antidepressants. Finally, we discuss the strengths and limitations of the model in Section 5.

## 2 Methods

We introduce a Bayesian hierarchical model for dose-response meta-analysis. We focus on a dichotomous outcome, although the models could easily accommodate continuous outcomes.

### 2.1 Notation

Table 1 summarizes the notation. Suppose there are $ns$ studies ($i = 1, ..., ns$) and each study has a number of doses $nd_i$ ($j = 1, ..., nd_i$). Each study reported an empirical estimate of the outcome at each dose level. The doses are denoted by $X_{ij}$ where the minimum dose $X_{i0}$ is set as the reference level (control group). The observed outcome is expressed as number of events out of total observed or relative treatment effects. The dose-specific number of events is $r_{ij}$ out of a total sample size $n_{ij}$. The estimated change in the outcome from the reference dose $X_{i0}$ to dose $X_{ij}$, summarized for the $n_{ij}$ participants, is indicated by $Y_{ij}$. $Y_{ij}$ can be log odds ratio (logOR), log risk ratio (logRR), log hazard ratio (logHR), or any relative treatment effect for continuous outcomes such as mean difference. Relative effects rather than number of events are commonly reported in the context of studying environmental exposures or other exposures examined in observational studies [16]. In this case, the relative effects $Y_{ij}$ are the estimates from multivariable models adjusted for possible confounding variables. The vector $\boldsymbol{Y}_i = (Y_{i1}, Y_{i2}, ..., Y_{i(nd_i-1)})$ comprises all relative effects, on a natural logarithmic scale, for study $i$.

### 2.2 Dose-response meta-analysis model

We propose a hierarchical two-level model. In the first level, the dose-response model is fitted within each study assuming either normal (normal dose-response model) or binomial likelihood (binomial dose-response model) for the observed data. In the second level, we synthesize the dose-response regression coefficients across studies. The hierarchical structure allows coefficients to borrow strength across studies, via the exchangeability assumption.

#### 2.2.1 Dose-response model within each study

Within each study $i$, a multivariate normal distribution is assumed for $\boldsymbol{Y}_i$

$$\boldsymbol{Y}_i \sim MVN(\boldsymbol{\Delta}_i, \boldsymbol{S}_i),$$

where the vector $\boldsymbol{\Delta_i} = (\delta_{i1}, \dots, \delta_{i(nd_i-1)})$ contains the *underlying* relative effects of dose $X_{ij}$ relative to dose $X_{i0}$. The $(nd_i\text{-}1) \times (nd_i\text{-}1)$ variance-covariance matrix $\boldsymbol{S_i}$ can be estimated assuming a multinomial distribution for the number of events per dose and using the delta-method for large sample sizes [17, 18]. For logOR, the elements of $\boldsymbol{S_i}$ are

$$\widehat{\sigma^2}_{ijm} = \begin{cases} 1/r_{i0} + 1/t_{i0}, & if\ j \neq m \\ 1/r_{ij} + 1/t_{ij} + 1/r_{i0} + 1/t_{i0}, & if\ j = m \end{cases},$$

where $t$ refers to the number of non-events and the zero index refers to the quantities in the reference dose. The formula above is suitable when the logORs are estimated from 2x2 tables. If the logORs originate from adjusted logistic models from observational studies, then a correction in the correlations between dose strata should be applied to $\boldsymbol{S_i}$, using the Longnecker and Greenland method [9, 10] or the approach suggested by Hamling [19].

If the data are from a randomized trial and the table of counts is available, it is straightforward to assume a binomial distribution of events

$$r_{ij} \sim Binom(p_{ij}, n_{ij}),$$

where $p_{ij}$ are the underlying probabilities of having an event in dose $j$ within study $i$. Then the underlying relative treatment effects are parametrised as

$$\lambda(p_{i0}) = u_i$$
$$\lambda(p_{ij}) = u_i + \delta_{ij},$$

with $\delta_{ij}$ defined as above. The function $\lambda$ is specified based on the effect size we want to estimate; for example, it is the logit function for logOR and the log function for logRR. The parameter $u_i$ is the log-odds of the event in the reference dose level.

Note that continuous outcome data can be accommodated if $\boldsymbol{Y_i}$ are mean differences or standardized mean differences. Alternatively, if the outcome is available for each dose level, the normal likelihood is used instead of the binomial, and $\delta_{ij}$ is parameterized as the mean difference or standardized mean difference.

### 2.2.2 *Dose-response functions*

The underlying relative effect $\delta_{ij}$ can be modelled as

$$\delta_{ij} = f(X_{ij}; X_{i0}; \boldsymbol{\beta_i}),$$

where $f$ is the dose-response function and $\boldsymbol{\beta_i}$ are the shape parameters that need to be estimated. Note that the $f$ function could also be any transformation, including linear, quadratic, cubic or fractional polynomials and resulting in $\boldsymbol{\beta_i} = (\beta_{ki})$ being a vector of length $p$ and $k = 1, 2, \dots, p$ [20]. The simplest case is to assume a linear ($f$ is the identity function) shape $p = 1$ where the statistical model needs to estimate only one parameter in study $i$; $\boldsymbol{\beta}_i = \beta_i$ and $\delta_{ij} = f(X_{ij}; X_{i0}; \beta_i) = \beta_i (X_{ij} - X_{i0})$. However, investigating dose-response relations underlying several studies may require non-linear models. A flexible choice is using restricted cubic splines [21]. With $m$ knots, there are $p = m - 1$ regression coefficients in $\boldsymbol{\beta}_i$ to be estimated. Setting $m = 3$, will result into $f$ consisting of $p = 2$ dose-transformations; $f_1$ is the identity function and $f_2$ the restricted cubic spline transformation [21] with coefficients $\boldsymbol{\beta}_i = (\beta_{1i}, \beta_{2i})$.

$$\delta_{ij} = \beta_{1i}\{f_1(X_{ij}) - f_1(X_{i0})\} + \beta_{2i}\{f_2(X_{ij}) - f_2(X_{i0})\}.$$

### 2.2.3 Synthesize dose-response functions across studies

In dose-response meta-analysis, the study-specific regression coefficients $\boldsymbol{\beta}_i = (\beta_{1i}, \beta_{2i}, ...)$ can then be synthesized. Random dose-response coefficients model assumes that the underlying study-specific coefficients $\boldsymbol{\beta}_i$ are normally distributed with mean $\boldsymbol{B} = (B_1, B_2, ... B_p)$ and variance-covariance matrix, $\boldsymbol{\Sigma}$, that is

$$\boldsymbol{\beta}_i \sim MVN(\boldsymbol{B}, \boldsymbol{\Sigma}).$$

This model acknowledges the presence of a distribution of true dose-response relationships underlying the studies and is capable of predicting study-specific curves by borrowing strength from their variation across studies. $\boldsymbol{\Sigma}$ is a $p \times p$ variance-covariance matrix with diagonal elements $\tau_p^2$ and in the off-diagonal there are the $p - 1$ covariances between the coefficients. To improve estimation, we will assume that $\tau_p^2 = \tau^2$ and we will explore whether the correlations $\rho$ in $\Sigma$ are different from zero. Note that this model assumes that the heterogeneity across the study-specific estimates is fully captured by heterogeneity in the dose-response shapes. For a model with a common dose-response coefficient we set $\boldsymbol{\beta}_i = \boldsymbol{B}$.

## 2.3 Dose-response meta-analysis model accounting for clustering in the exposure

Consider an exposure (or drug) variable that can take on different values. For example, daily intake of omega 3 fatty acids in relation to risk of cardiovascular events, possibly accounting for the different assessment of omega 3 (food supplements versus diet with fish and nuts). The differences between these two dose-response curves can be modelled by inserting type-specific regression coefficients $\boldsymbol{\beta}_i^c = (\beta_{1i}^c, \beta_{2i}^c, ... \beta_{pi}^c)$, where

$$c = \{1: \text{food supplements}, 2: \text{diet with fish and nuts}\}.$$

Overall, for a random of exposure clusters $c = 1, 2, \ldots, C$ the relative effects are mapped to the transformed dose as

$$\delta_{ij} = f(X_{ij}; X_{i0}; \boldsymbol{\beta}_i^c).$$

Next, $\boldsymbol{\beta}_i^c$, the vectors of coefficients from study $i$ examining the same cluster of exposures, are synthesised using a multivariate normal distribution with a common mean $\boldsymbol{B}^c = (B_1^c, B_2^c, \ldots B_p^c)$ and variance-covariance matrix $\boldsymbol{\Sigma}^{within}$; that is a $p \times p$ matrix with diagonal $\tau_{within}^2$ and on the off-diagonal the $p-1$ covariances between the coefficients

$$\boldsymbol{\beta}_i^c \sim MVN(\boldsymbol{B}^c, \boldsymbol{\Sigma}^{within}).$$

At the next step, the cluster-specific dose-response associations $\boldsymbol{B}^c$ are synthesised across the $C$ clusters. Again, a multivariate normal distribution with mean vector $\boldsymbol{B}$ and variance-covariance matrix $\boldsymbol{\Sigma}^{between}$ is employed. $\boldsymbol{\Sigma}^{between}$ has the same dimension as $\boldsymbol{\Sigma}^{within}$ and in the diagonal the parameter $\tau_{between}^2$ measures the heterogeneity between the clusters

$$\boldsymbol{B}^c \sim MVN(\boldsymbol{B}, \boldsymbol{\Sigma}^{between})$$

## 2.4 Predictions for the absolute response to a dose

Predictions are easy to make within a Bayesian model as the total uncertainty in the parameters is propagated in the final predictions. Assume there is a natural reference dose, such as a dose zero or no-exposure. The observations $r_{i0}, n_{i0}$ from the zero dose levels can be parametrised to estimate an average summary response to zero-dose $R_0$

$$r_{i0} \sim Binom(p_{i0}, n_{i0}),$$
$$\lambda(p_{i0}) \sim N(R_0, \sigma_0^2).$$

Then, the estimate $R_0$ (measured on the log or logit probability scale) can be combined with $\boldsymbol{B}$ to obtain the absolute response to any given dose level $X_j$

$$\lambda^{-1}\{B_1\{f_1(X_j) - f_1(X_0)\} + B_2\{f_2(X_j) - f_2(X_0)\} + R_0\}$$

## 2.5 Bayesian estimation

We will use *Markov chain Monte Carlo* (MCMC) techniques to estimate all parameters in a Bayesian setting. An approximate non-informative prior distribution is chosen for the coefficients and the baseline effects $u_{i0} = \text{logit}(p_{i0})$ in the binomial model

$$B_k \sim N(0, 10^3)$$
$$u_{i0} \sim N(0, 10^3).$$

Given that both in the simulations and in the example our outcome is dichotomous and measured on the natural log scale, we place a half-normal prior to the heterogeneity parameter

$$\tau \sim N_0(0,1)$$

For correlations $\rho$ in the off-diagonal of the variance-covariances matrices, we use a uniform prior:

$$\rho \sim Unif(-1,1).$$

All Bayesian models are implemented in JAGS within R [22, 23]. The codes can be found in GitHub at https://github.com/htx-r/DoseResponsePMA. To obtain the spline transformations, we use the *rcs* function from the *rms* package [24]. To evaluate the convergence of the models we employed various diagnostic tools for MCMC included in the *coda* package [25]. We explored convergence plots for the MCMC (histograms, trace plots, Geweke plot and Gelman-Rubin plot) and relevant statistics (Raftery and Lewis statistic and Heidelberger and Welch test) [26].

## 3 Simulations study

We aim to investigate the agreement between the estimations of the dose-response meta-analysis curve under our two Bayesian models, assuming random-effects for the coefficients, and the frequentist one-stage model [27]. The codes are available in GitHub.

### 3.1 Simulation design

We assumed a true restricted cubic spline dose-response relationship with 3 knots at fixed percentiles (25th, 50th, and 75th) of the dose. We modelled the logOR and the logRR.

For 40 clinical trials, we simulated study-level aggregated data. For each study, we simulated two non-zero doses from uniform distribution $X_{ij} \sim U(1,10)$ and assumed each study reported one zero dose. The study-specific coefficients $\beta_{i1}$ and $\beta_{i2}$ are generated independently from univariate normal distribution with means $B_1$ and $B_2$, respectively, and common heterogeneity $\tau$. We chose true coefficient values that cover a reasonable range for ORs (0.3 to 5) and we considered different dose-response shapes (see Appendix Figure 1). We introduced between-study heterogeneity, $\tau = 0.001, 0.01$. The assumed mean and heterogeneity values result in eight scenarios, as shown in Table 2.

Using $\beta_{1i}, \beta_{2i}$ and $X_{ij}$, we calculated the underlying treatment effect $\delta_{ij} = \log OR_{ij}$. To improve computing time, we assumed that the two shape coefficients $\beta_{1i}$ and $\beta_{2i}$ are unrelated ($\rho = 0$). Per dose, the observed number of events $r_{ij}$ are generated from binomial distributions with probability $p_{ij}$ and sample size $n_{ij}$. The event rate in the zero-dose group $p_0$ is set to 0.1. The underlying event rate at dose $j$ is $p_{ij} = \exp(\delta_{ij}) \times p_0$. The sample size per dose is generated from a uniform distribution $n_{ij} \sim Unif(180, 220)$. In this way, the number of events and sample size per dose for each study are generated and used as input for the Bayesian binomial model. Using these counts, we then estimate $\log \widehat{OR}_{ij}$ and their standard errors to use as inputs for the Bayesian normal and frequentist models [27].

Following the same steps as logOR above, we simulated the dataset expressing the underlying treatment effect, instead, in terms of risk ratio; $\delta_{ij} = \log RR_{ij}$. The additional concern, particularly for RR, that we need to confine probabilities within 0 and 1. Therefore, we inserted, $maxRR = \exp((B_k + 2\tau) \times \max(f(X_{ij})))$ then we set $p_0 = 0.5/maxRR$. Along with that, we restrict the values of both $p_0$ and $p_1$; $0.05 < p_0 < 0.95$ and $p_1 < 0.97$, to avoid numerical problems that emerge near the boundaries.

The Bayesian models were estimated using $1 \times 10^5$ iterations with three chains, with a burn-in of $1 \times 10^4$ and a thinning of one. Given that the simulated data was produced assuming $\rho = 0$, we did not use bivariate distributions but two independent distributions for $\beta_{1i}$ and $\beta_{2i}$. Each scenario was studied in 1000 simulations. We used the *dosresmeta* command to fit the frequentist model [28].

For each method, we estimated the mean bias in the regression coefficients $B_1$ and $B_2$ and $\tau$ as the difference between the true coefficient and the corresponding mean estimated value. We computed the mean squared error (MSE) as the sum of the squared bias and the variance of the estimates to quantify the variation in sample estimates. As graphical output is difficult to monitor in a simulation study, the convergence of the MCMC was quantified here only by computing the Gelman statistics $\sqrt{\hat{R}}$; when $\sqrt{\hat{R}} \approx 1$ the MCMC converges. Additionally, we report the coverage for each estimate as the proportion of credible intervals that captured the true value. We computed the power to detect $B_k \neq 0$ when the estimated credible interval does not include zero and the mean of the coefficients' standard error (SE2mean). Finally, we report the Monte Carlo standard error (MCse) to quantify the uncertainty of all the quantities presented above. We present the results from OR for bias and MSE in the main text whereas the remaining results are presented in the Appendix.

## 3.2 Simulation Results

Table 2 presents the results from the eight scenarios for logORs using splines. Figure 1 shows the average estimated curves for scenarios 2 to 4 (results from scenarios 6 to 8 provide similar conclusions to those in Figure 1; scenarios 1 and 5 refer to no dose-response association and are not presented in the figure).

The three estimated dose-response lines are indistinguishable in Figure 1 and all three models perform very well and give practically identical results (Table *2*). The binomial Bayesian model has a slightly lower bias in the coefficients than the normal Bayesian and the frequentist approach in all scenarios. The spline coefficients $B_2$ exhibit more bias and are less accurate than those of $B_1$. For both binomial and normal Bayesian models, larger heterogeneity $\tau = 0.01$ resulted in considerably less bias than when $\tau = 0.001$.

Additional results from the simulations are presented in Appendix Table 1-3. The coverage of all estimates exceeds 90%. The power to detect a nonzero linear coefficient $B_1$ ranges between 85% and 93% when $B_1 = 0.04$ and 100% for $B_1 = 0.2$. The power to detect a non-linear association, ranges between 20% and 28% when $B_2 = 0.03$ and is 100% when $B_2 = -0.2$. Whereas, the power to detect $\tau$ is very low. The MCMC converged in all simulations as $\sqrt{\hat{R}} < 1.015$. Finally, the largest MCse of bias is $9 \times 10^{-4}$.

The results for logRR are presented in Appendix Figure 11 and Appendix Table 4-6. The results of logRR are actually agree with the ones based on logOR. The binomial Bayesian model has a slightly less bias and MSE in the coefficients than the normal Bayesian model and the one-stage approach in most scenarios. Likewise, the binomial model conveys better coverage than the normal Bayesian and the one-stage approach. Regarding power, all approaches perform well and the three dose-response curves are identical in Appendix Figure 11. The coverage in binomial model is much higher than the normal model, whereas the opposite is obtained regarding the power. Convergence was good as $\sqrt{\hat{R}} < 1.05$ in all cases. The largest MCse of bias is $9 \times 10^{-4}$.

## 4 Dose-response function for antidepressants in major depression

We illustrate the methods by synthesizing the dose-response association reported in 60 randomized controlled trials (145 arms, 15,174 participants) examining the efficacy and tolerability of various doses of SSRI antidepressant drugs [29]. Using a previously validated formula, we first transformed the dosages of the different antidepressants into fluoxetine-equivalents [29]. The response to antidepressant is defined as 50% reduction in symptoms. The data and analysis are also available in the GitHub directory. We estimated the dose-response relationship using restricted cubic spline with three knots placed at fixed percentiles of the dose: 10, 20, and 50 mg/day.

The results are displayed in Table 3 and the dose-response curves based on the three approaches are shown in Figure 2a. The estimated correlation $\rho$ has a substantial uncertainty. The two Bayesian models agree to a large extent with the frequentist approach in the estimated linear and spline coefficients and in the precision of the estimations, as shown in results in Table 3. There are immaterial differences between the frequentist and the Bayesian models in the estimation of heterogeneity and correlation $\rho$; the latter is estimated with large uncertainty in Bayesian models.

In Figure 2b we present the absolute response at each dose level between 1 to 80 mg/day estimated using the binomial Bayesian model and the approach presented in section 2.4. The uncertainty in the dose-response curve is smaller for smaller doses and gets wider for higher doses, as less data is available. The response in the placebo arm was estimated at 37.6% (blue line in Figure 2b) [30].

We also fit the clustered dose-response model where studies have first being synthesised within drug and then across drugs using the binomial likelihood. The coefficients $B_1, B_2$ were very similar to those estimated from the model that ignores clusters (see Table 3). The within-drug variance $\tau_{within}$ was estimated 0.0076, a bit smaller than the total heterogeneity from the binomial model ($\tau = 0.0087$). There were some differences between the eight drugs as indicated from the $\tau_{between} = 0.0050$. However, the dose-response shape is practically identical to that of the model that ignores the drug clustering. Finally, the within and between cluster correlations are estimated with large uncertainty like in all models.

We examined the convergence of MCMC for all Bayesian models. Overall, convergence is achieved for the all estimated parameters of the three models, see Appendix Table 8-13 and Appendix Figure 13-25.

## 5 Discussion

In this paper, we present a hierarchical dose-response meta-analysis model in a Bayesian framework. At the first level, the dose-response relationship is fitted within each study. Then the curves are combined to get the average dose-response. An additional pooling level can be added, if there are different clusters of exposure or drugs. The exact likelihood of the outcome (binomial or normal) can be employed if arm-level data is available. Alternatively, the observed relative contrast between the study-specific lowest dose and each subsequent dose level are assumed to follow a normal likelihood. The model accounts for the covariance of the effects of multiple doses and the variability in the dose-response association between studies. We showed that the model using the binomial likelihood and normal likelihood performs as well as the frequentist one-stage model and that bias in the coefficients is slightly smaller for the binomial Bayesian model.

Among the limitations common to all Bayesian approaches, two are particularly challenging for our model [15, 31]. First, for some scenarios, the estimation can be sensitive to the prior choice [32]. In these cases, sensitivity analysis is recommended with either different prior distributions or by varying the characteristics (hyperparameters) of the specific prior distribution. Second, time-consuming, intensive computation may be required until MCMC convergence is achieved. In this context, we emphasize the importance of investigating the convergence of MCMC using CODA approaches (e.g. as those presented in appendix). Furthermore, the usual challenges of dose-response meta-analysis apply, including ambiguity in the categorisation of the exposure, the reporting of different categories by different studies or of open-ended categories. These issues are discussed in detail elsewhere [16].

A strength of our Bayesian approach is its flexibility. We were able to evaluate whether studies that examine the same drug are more similar than studies examining different drugs by using an extension of our model that adds a layer in the hierarchy according to the specific kind of antidepressant that was studied. We were also able to estimate the absolute response to each dose [14, 15]. Such estimates can also be obtained in a frequentist setting by using best linear unbiased prediction (BULPs) in mixed models, [33, 34]. However, the process is easier in a Bayesian framework, which also allows the use of external data to estimate the outcome at zero dose. The approach will be particularly valuable in the context of policy- and decision-making where the absolute event rates play a more important role than the relative treatment effects.

The hierarchical structure of the model allows the borrowing of strength across studies [12]. Studies that report only one dose-specific effect can thus be included and a nonlinear dose-response model fitted. This is also possible in a frequentist setting using the one-stage approach, however, our model can be extended to separate between the heterogeneity due to variability in dose-response shape and residual between-study heterogeneity. The latter can be explored by including covariates that may explain this residual variability; that could lead into a dose-response meta-regression. Our model could also be extended to multiple treatments, thus offering an alternative to published network meta-analysis models [35], or it could be used to model simultaneously several outcomes with similar dose-response shapes. Another potential extension, which we have implemented in our paper, is accounting for cluster of the exposure in estimating the dose-response shape. Finally, external knowledge can be incorporated, for example, evidence from observational studies. The use of observational data will be particularly relevant when assessing long term outcomes, as the majority of RCTs, in psychiatry and elsewhere, are of relatively short duration [17].

In conclusion, the proposed binomial or normal Bayesian dose-response model provides a viable alternative to the existing mixed one-stage model in a frequentist setting. Researchers can take advantage of the high flexibility of the model to address complex problems and multiple sources of heterogeneity.

Table 1 Notation in aggregated-level data in dose-response meta-analysis

| | |
|---|---|
| $i = 1, \ldots, ns$ | Study id |
| $j = 1, \ldots, nd_i$ | Dose levels in study $i$ |
| $X_{ij}$ | Dose level $j$ in study $i$ |
| $X_{i0}$ | Reference dose in study $i$ |
| $r_{ij}$ | Number of events in dose $j$ within study $i$ |
| $n_{ij}$ | Sample size in dose $j$ within study $i$ |
| $Y_{ij}$ | Within study $i$, the relative effect (on a ln-scale) of dose $j$ contrasted to the effect in the reference dose ($X_{i0}$) e.g. log odds ratio |
| $\boldsymbol{Y_i} = \left(Y_{i1}, Y_{i2}, \ldots, Y_{i(nd_i-1)}\right)$ | Vector of all dose-specific (ln) relative effects in study $i$ |
| $k = 1, \ldots p$ | Number of dose transformations associated with the dose-response shape. For a linear shape $p = 1$ and for quadratic and restricted cubic splines $p = 2$ |
| $c = 1, 2, \ldots, C$ | Exposure clusters |

Table 2. Simulations scenarios for a spline dose-response association assuming random effects for $B_1$, $B_2$. We assume 40 trials reporting aggregated-level data with three dose-levels each. The bias and MSE are reported for linear coefficient, spline coefficient and their common heterogeneity (a) $B_1$ (b) $B_2$ (c) $\tau$, respectively. **Bias and MSE are divided by $10^3$.**

(a) Estimate $B_1$

| | True values | | | Binomial Bayesian | | Normal Bayesian | | One-stage (frequentist) | |
|---|---|---|---|---|---|---|---|---|---|
| Scenario | $\tau$ | $B_1$ | $B_2$ | bias | MSE | bias | MSE | bias | MSE |
| S1 | 0.001 | 0 | 0 | 0.1 | 0.2 | 5.5 | 0.2 | 4.8 | 0.2 |
| S2 | 0.001 | 0.04 | 0 | 0.2 | 0.2 | 4.9 | 0.2 | 4.3 | 0.2 |
| S3 | 0.001 | 0.1 | 0.03 | 0.4 | 0.2 | 4.5 | 0.2 | 4 | 0.2 |
| S4 | 0.001 | 0.2 | -0.2 | 1 | 0.1 | 3.8 | 0.2 | 4.1 | 0.2 |
| S5 | 0.01 | 0 | 0 | 0.4 | 0.2 | 5.8 | 0.2 | 5.2 | 0.2 |
| S6 | 0.01 | 0.04 | 0 | 0.8 | 0.2 | 5.6 | 0.2 | 5 | 0.2 |
| S7 | 0.01 | 0.1 | 0.03 | 0.9 | 0.2 | 4.8 | 0.2 | 4.4 | 0.2 |
| S8 | 0.01 | 0.2 | -0.2 | 2 | 0.1 | 4.6 | 0.2 | 4.9 | 0.2 |

(b) Estimate $B_2$

| | True values | | | Binomial Bayesian | | Normal Bayesian | | One-stage (frequentist) | |
|---|---|---|---|---|---|---|---|---|---|
| Scenario | $\tau$ | $B_1$ | $B_2$ | bias | MSE | bias | MSE | bias | MSE |
| S1 | 0.001 | 0 | 0 | -0.3 | 0.8 | -7.1 | 0.8 | -5 | 0.8 |
| S2 | 0.001 | 0.04 | 0 | -1.2 | 0.7 | -7.5 | 0.7 | -6.1 | 0.7 |
| S3 | 0.001 | 0.1 | 0.03 | 0.7 | 0.6 | -5.7 | 0.6 | -4.3 | 0.6 |
| S4 | 0.001 | 0.2 | -0.2 | -0.9 | 0.5 | -4.2 | 0.5 | -4.3 | 0.5 |
| S5 | 0.01 | 0 | 0 | -0.8 | 0.9 | -7.8 | 0.9 | -6 | 0.9 |
| S6 | 0.01 | 0.04 | 0 | -0.4 | 0.7 | -6.4 | 0.7 | -4.8 | 0.7 |
| S7 | 0.01 | 0.1 | 0.03 | 0 | 0.6 | -6.3 | 0.6 | -4.8 | 0.6 |
| S8 | 0.01 | 0.2 | -0.2 | -2.6 | 0.5 | -5.7 | 0.6 | -5.9 | 0.6 |

(c) Estimate $\tau$

| | True values | | | Binomial Bayesian | | Normal Bayesian | |
|---|---|---|---|---|---|---|---|
| Scenario | $\tau$ | $B_1$ | $B_2$ | bias | MSE | bias | MSE |
| S1 | 0.001 | 0 | 0 | 12.4 | 0.2 | 13 | 0.2 |
| S2 | 0.001 | 0.04 | 0 | 11.8 | 0.2 | 12.6 | 0.2 |
| S3 | 0.001 | 0.1 | 0.03 | 10.5 | 0.1 | 11.6 | 0.2 |
| S4 | 0.001 | 0.2 | -0.2 | 10.8 | 0.1 | 11.7 | 0.2 |
| S5 | 0.01 | 0 | 0 | 5.1 | 0.1 | 5.5 | 0.1 |
| S6 | 0.01 | 0.04 | 0 | 4.4 | 0.1 | 5.1 | 0.1 |
| S7 | 0.01 | 0.1 | 0.03 | 3.3 | 0 | 4.1 | 0.1 |
| S8 | 0.01 | 0.2 | -0.2 | 3.7 | 0 | 4.7 | 0.1 |

Table 3 Dose-response between antidepressants and response to drug. The model is fitted with restricted cubic splines and assuming random dose-response coefficients. Dose is measured as fluoxetine-equivalent in mg/day.

| | Binomial Bayesian | | Normal Bayesian | | one-stage (frequentist) | | Binomial Bayesian with drug clusters | |
|---|---|---|---|---|---|---|---|---|
| | Mean | SD | Mean | SD | Mean | SE | Mean | SD |
| $B_1$ | 0.0214 | 0.0024 | 0.0210 | 0.0037 | 0.0209 | 0.0025 | 0.0213 | 0.0036 |
| $B_2$ | -0.0397 | 0.0070 | -0.0396 | 0.0085 | -0.0376 | 0.0060 | -0.0387 | 0.0079 |
| $\tau$ | 0.0087 | 0.0028 | 0.0072 | 0.0031 | $\tau_1 = 0.0103$ $\tau_2 = 0.0115$ | - | $\tau_{within} = 0.0076$ $\tau_{between} = 0.0050$ | 0.0028 0.0040 |
| $\rho$ | -0.4782 | 0.4952 | -0.2488 | 0.5652 | -1 | - | $\rho_{within} = -0.3611$ $\rho_{between} = -0.1064$ | 0.5153 0.5508 |

## *Figures captions*

Figure 1 Dose-response associations corresponding to scenarios 2-4 are in upper three panels. The lower three panels are a snapshot in the lager dose range 8-10 to investigate the slight differences between the three approaches.

Figure 2 (a) The relative dose-response associations estimated with the three approaches; binomial Bayesian, normal Bayesian and one-stage (frequentist) approaches. Analyses based on 60 randomized clinical trials of antidepressant drugs. (b) The absolute response to antidepressants at each dose level over a range of 1 to 80 mg/day, estimated using the binomial Bayesian model. The dashed lines represent the boundaries of the credible region around the absolute dose-response curve. The red lines are the estimated placebo response (solid line) and the 95% boundaries of the credible interval (dashed lines).

# Appendix

## *Tables captions*

Appendix Table 1 Assuming odds ratio (OR) as a measure of the treatment effect, 8 scenarios are simulated for a spline dose-response association with random effects coefficients. We assume 40 trials reporting aggregated-level data with three dose-levels each. The results for the linear coefficient $B_1$.

Appendix Table 2 Assuming odds ratio (OR) as a measure of the treatment effect, 8 scenarios are simulated for a spline dose-response association with random effects coefficients. We assume 40 trials reporting aggregated-level data with three dose-levels each. The results for the spline coefficient $B_2$.

Appendix Table 3 Assuming odds ratio (OR) as a measure of the treatment effect, 8 scenarios are simulated for a spline dose-response association with random effects coefficients. We assume 40 trials reporting aggregated-level data with three dose-levels each. The results for the common heterogeneity $\tau$.

Appendix Table 4 Assuming risk ratio (RR) as a measure of the treatment effect, 8 scenarios are simulated for a spline dose-response association with random effects coefficients. We assume 40 trials reporting aggregated-level data with three dose-levels each. The results for the linear coefficient $B_1$.

Appendix Table 5 Assuming risk ratio (RR) as a measure of the treatment effect, 8 scenarios are simulated for a spline dose-response association with random effects coefficients. We assume 40 trials reporting aggregated-level data with three dose-levels each. The results for the spline coefficient $B_2$.

Appendix Table 6 Assuming risk ratio (RR) as a measure of the treatment effect, 8 scenarios are simulated for a spline dose-response association with random effects coefficients. We assume 40 trials reporting aggregated-level data with three dose-levels each. The results for the common heterogeneity $\tau$.

Appendix Table 7 Results of the three approaches regarding the estimation of linear and spline coefficients $B_1$ and $B_2$, respectively, in addition to their common heterogeneity $\tau$. These results are based on simulated antidepressant dataset from restricted cubic spline dose-response meta-analysis model, the coefficients are set as the frequentist estimation that are displayed in Table 3.

Appendix Table 8 For binomial Bayesian model, the estimated number of burn-in, number of iterations and I factor on each chain is presented.

Appendix Table 9 For normal Bayesian model, the estimated number of burn-in, number of iterations and I factor on each chain is presented.

Appendix Table 10 For binomial Bayesian model with drug-specific class, the estimated number of burn-in, number of iterations and I factor on each chain is presented.

Appendix Table 11 For binomial Bayesian model, the p-value of the stationarity test and halfwidth test and the estimated posterior mean of the stationary part of chain is displayed.

Appendix Table 12 For normal Bayesian model, the p-value of the stationarity test and halfwidth test and the estimated posterior mean of the stationary part of chain is displayed.

Appendix Table 13 For binomial Bayesian model with drug-specific class, the p-value of the stationarity test and halfwidth test and the estimated posterior mean of the stationary part of chain is displayed.

## *Figures captions*

Appendix Figure 1 The underlying dose-response curve in simulations with its boundaries as true curve ± 2*$\tau$ (dotted lines) assuming small (red) and large (blue) values for heterogeneity $\tau$.

Appendix Figure 2 Dose-response associations corresponding to scenarios 2-4 are in upper three panels. The lower three panels are a snapshot in the lager dose range 8-10 to investigate the slight differences between the three approaches.

Appendix Figure 3 Histogram for $\hat{B}_1$ in a simulation study based on odds ratio (OR) for the binomial dose-response meta-analysis model of restricted cubic spline in various scenarios for true $B_2$= (a) 0 (b) 0 (c) 0.03 (d) -0.2 and true $B_1$= (a) 0 (b) 0.04 (c) 0.1 (d)0.2 (green line) where in the first and the second columns the true heterogeneity is set as $\tau = 0.001$ and $\tau = 0.01$, respectively.

Appendix Figure 4 Histogram for $\hat{B}_2$ in a simulation study based on odds ratio (OR) for the binomial dose-response meta-analysis model of restricted cubic spline in various scenarios for true $B_2$= (a) 0 (b) 0 (c) 0.03 (d) -0.2 (green line) and true $B_1$= (a) 0 (b) 0.04 (c) 0.1 (d)0.2 where in the first and the second columns the true heterogeneity is set as $\tau = 0.001$ and $\tau = 0.01$, respectively.

Appendix Figure 5 Histogram for $\hat{B}_1$ in a simulation study based on odds ratio (OR) for the normal dose-response meta-analysis model of restricted cubic spline in various scenarios for true $B_2$= (a) 0 (b) 0 (c) 0.03 (d) -0.2 and true $B_1$= (a) 0 (b) 0.04 (c) 0.1 (d)0.2 (green line) where in the first and the second columns the true heterogeneity is set as $\tau = 0.001$ and $\tau = 0.01$, respectively.

Appendix Figure 6 Histogram for $\hat{B}_2$ in a simulation study based on odds ratio (OR) for the normal dose-response meta-analysis model of restricted cubic spline in various scenarios for true $B_2$= (a) 0 (b) 0 (c) 0.03 (d) -0.2 (green line) and true $B_1$= (a) 0 (b) 0.04 (c) 0.1 (d)0.2 where in the first and the second columns the true heterogeneity is set as $\tau = 0.001$ and $\tau = 0.01$, respectively.

Appendix Figure 7 Histogram for $\hat{\tau}$ in a simulation study based on odds ratio (OR) for the binomial dose-response meta-analysis model of restricted cubic spline in various scenarios for true $B_2$= (a) 0 (b) 0 (c) 0.03 (d) -0.2 and true $B_1$= (a) 0 (b) 0.04 (c) 0.1 (d)0.2 where in the first and the second columns the true heterogeneity is set as $\tau = 0.001$ and $\tau = 0.01$ (green line), respectively.

Appendix Figure 8 Histogram for $\hat{\tau}$ in a simulation study based on odds ratio (OR) for the normal dose-response meta-analysis model of restricted cubic spline in various scenarios for true $B_2$= (a) 0 (b) 0 (c) 0.03 (d) -0.2 and true $B_1$= (a) 0 (b) 0.04 (c) 0.1 (d)0.2 where in the first and the second columns the true heterogeneity is set as $\tau = 0.001$ and $\tau = 0.01$ (green line), respectively.

Appendix Figure 9 Histogram for $\hat{B}_1$ in a simulation study based on odds ratio (OR) for the one-stage (frequentist) dose-response meta-analysis model of restricted cubic spline in various scenarios for true $B_2$= (a) 0 (b) 0 (c) 0.03 (d) -0.2 and true $B_1$= (a) 0 (b) 0.04 (c) 0.1 (d)0.2 (green line) where in the first and the second columns the true heterogeneity is set as $\tau = 0.001$ and $\tau = 0.01$, respectively.

Appendix Figure 10 Histogram for $\hat{B}_2$ in a simulation study based on odds ratio (OR) for the one-stage (frequentist) dose-response meta-analysis model of restricted cubic spline in various scenarios for true $B_2$= (a) 0 (b) 0 (c) 0.03 (d) -0.2 (green line) and true $B_1$= (a) 0 (b) 0.04 (c) 0.1 (d)0.2 where in the first and the second columns the true heterogeneity is set as $\tau = 0.001$ and $\tau = 0.01$, respectively.

Appendix Figure 11 Dose-response associations corresponding to scenarios 2-4 are in the above three panels. The lower three panels are a snapshot in the lager dose range 8-10 to identify the slight differences between the three approaches. In simulations and model, risk ratio (RR) has been used as a measure of the treatment effect.

Appendix Figure 12 The estimated dose-response curves of the 60 randomized clinical trials that studied the effectiveness of antidepressant drugs.

Appendix Figure 13 The distribution of $\hat{B}_1$, $\hat{B}_2$, $\hat{\tau}$ and $\hat{\rho}$ of the binomial Bayesian model.

Appendix Figure 14 The distribution of $\hat{B}_1$, $\hat{B}_2$, $\hat{\tau}$ and $\hat{\rho}$ of the normal Bayesian model.

Appendix Figure 15 The distribution of $\hat{B}_1$, $\hat{B}_2$, $\hat{\tau}_{within}$, $\hat{\tau}_{between}$, $\hat{\rho}_{within}$ and $\hat{\rho}_{between}$ of the binomial Bayesian model with drug-specific class.

Appendix Figure 16 The trace plot of $\hat{B}_1$, $\hat{B}_2$, $\hat{\tau}$ and $\hat{\rho}$ of the binomial Bayesian model.

Appendix Figure 17 The trace plot $\hat{B}_1$, $\hat{B}_2$, $\hat{\tau}$ and $\hat{\rho}$ of the normal Bayesian model.

Appendix Figure 18 The trace plot of $\hat{B}_1$, $\hat{B}_2$, $\hat{\tau}_{within}$, $\hat{\tau}_{between}$, $\hat{\rho}_{within}$ and $\hat{\rho}_{between}$ of the binomial Bayesian model with drug-specific class.

Appendix Figure 19 The Geweke plot for each chain of $\hat{B}_1$, $\hat{B}_2$, $\hat{\tau}$ and $\hat{\rho}$ of the binomial Bayesian model.

Appendix Figure 20 The Geweke plot for each chain of $\hat{B}_1$, $\hat{B}_2$, $\hat{\tau}$ and $\hat{\rho}$ of the normal Bayesian model.

Appendix Figure 21 The Geweke plot for each chain of $\hat{B}_1$, $\hat{B}_2$ and $\hat{\tau}_{within}$ of the binomial Bayesian model with drug-specific class.

Appendix Figure 22 The Geweke plot for each chain $\hat{\tau}_{between}$, $\hat{\rho}_{within}$ and $\hat{\rho}_{between}$ of the binomial Bayesian model with drug-specific class.

Appendix Figure 23 For binomial Bayesian model, Gelman-Rubin plot of the shrink factor $\sqrt{\hat{R}}$ over the last iterations is displayed $\hat{B}_1$, $\hat{B}_2$, $\hat{\tau}$ and $\hat{\rho}$.

Appendix Figure 24 For normal Bayesian model, Gelman-Rubin plot of the shrink factor $\sqrt{\hat{R}}$ over the last iterations is displayed $\hat{B}_1$, $\hat{B}_2$, $\hat{\tau}$ and $\hat{\rho}$.

Appendix Figure 25 For binomial Bayesian model with drug-specific class, Gelman-Rubin plot of the shrink factor $\sqrt{\hat{R}}$ over the iterations is displayed for $\hat{B}_1$, $\hat{B}_2$, $\hat{\tau}_{within}$, $\hat{\tau}_{between}$, $\hat{\rho}_{within}$ and $\hat{\rho}_{between}$.